**Dynamics of Coupled Topological – Collective Modes of a Bose-Einstein Condensate**

Aranya B. Bhattacherjee
Department of Physics, Atma Ram Sanatan Dharma College
(University of Delhi, South Campus),
Dhaula Kuan, New Delhi- 110021
India,
Email: abhattac@bol.net.in

**Abstract:** We develop the theory of the resonant formation of coupled topological-collective coherent modes in the presence of a quantized trap and classical external field. The coupling between the topological and the collective modes can be linear as well as non-linear depending upon the tuning of the external extremely low frequency electromagnetic field. This tuning depends on the trap frequency and the energy level separation between the ground and the first excited topological coherent mode modified by the two body atomic collisions.

**Introduction**

One of the most fascinating predictions of quantum statistical mechanics is that of a phase transition in an ideal gas of identical bosons when the thermal de Broglie wavelength exceeds the mean spacing between particles. Under such conditions, bosons are stimulated by the presence of other bosons in the lowest energy state to occupy that state as well, resulting in a macroscopic occupation of a single quantum state. This transition is termed Bose-Einstein Condensation (BEC) and the condensate that forms constitutes a macroscopic quantum-mechanical object. In the mean field regime, the Bose system is adequately described by the Gross-Pitaevskii equation. The stationary solutions of the Gross-Pitaevskii equation are called coherent modes. Thus BEC of trapped atoms can be understood as the condensation to the ground coherent state. These coherent modes are also termed as topological because the solutions corresponding to different spectral levels have principally different spatial dependences. The ground coherent state is an equilibrium statistical system. It was proposed that it is possible to create non-ground state condensates of bose atoms i.e. higher topological states [1]. Clearly, such a BEC will be a non-equilibrium system. If an additional field is switched on, with a frequency in resonance with the transition frequency between two coherent energy levels, then higher topological modes can be excited. For instance, a rotating field using multiple lasers can excite vortices [2-5], which are topological states.

There is other numerous examples of topological states, which have been studied, in the recent past [6,7]. Various properties of these coherent topological modes such as temporal behaviour, collective excitations have been investigated [8-10]. In a recent interesting work, it was observed that a resonant excitation does not guarantee the creation of higher topological modes [11]. To excite higher topological modes, in addition to a resonance transition, one needs either a harmonic confining potential or a nonlinear spatial interaction between the bose atoms and the additional time-dependent field. In our previous work, we had developed a quantum theory of the formation of topological coherent modes, in a BEC. We had demonstrated that the two body atomic interactions modifies the Rabi frequency and the well known phenomena of collapse and revival of Rabi oscillations is observed if the granular structure of the external perturbing field is taken into account [12]. The aim of the present paper is to develop the theory of the resonant formation of topological coherent modes in the presence of a quantized trap and classical external field. In other words, the center of mass motion of the BEC is quantized.

**The Model**

We take as our starting point a gas of two-level atoms that have experienced Bose-Einstein condensation, contained within a magnetic trap. The basic level scheme is shown in Fig.1.Under equilibrium conditions, the Bose-condensate is always the ground single particle state. In order to excite higher topological modes, an external extremely low frequency (ELF, 10-200 Hz) is directed along the x-axis. The ELF electromagnetic wave is taken to be classical. It is to be noted that the transition frequency between two topological coherent energy levels is of the order of trap frequencies (10-200 Hz). The trap is approximated as a harmonic oscillator of frequency $v$. The value of this trap frequency is determined by the various trap parameters. The operator $\hat{x}$ gives the position of the center-of-mass of the condensate, which we shall write below in terms of operators $\hat{a}$ and $\hat{a}^+$. These are now taken to be operators that annihilate and create quanta of the collective oscillator modes of the center-of-mass of the BEC. This basically corresponds to translation of the BEC cloud without change in its internal structure. The Hamiltonian for our system is given by

$$H = H_{atom} + H_v + H_{atom-field} + H_{atom-atom} \qquad (1)$$

Here $H_{atom}$ atom is the free evolution of the atomic-field. $H_v$ describes the collective-elementary excitation of the center-of-mass. $H_{atom-field}$ describes the classical coupling between the atomic field and the external ELF electromagnetic field. $H_{atom-atom}$ contains the two-body s-wave scattering collisions between atoms. The free atomic Hamiltonian is given by

$$H_{atom} = \int d^3\vec{r} \begin{bmatrix} \psi_g^+(\vec{r})\left(-\dfrac{\hbar^2\vec{\nabla}^2}{2m} + V_g(\vec{r})\right)\psi_g(\vec{r}) + \\ \psi_e^+(\vec{r})\left(-\dfrac{\hbar^2\vec{\nabla}^2}{2m} + V_e(\vec{r})\right)\psi_e(\vec{r}) \end{bmatrix} \qquad (2)$$

Where $m$ is the atomic mass. $V_g(\vec{r})$ and $V_e(\vec{r})$ are the trap potentials of the ground and the excited states. $H_v$ is written as

$$H_v = \hbar v \left(\hat{a}^+\hat{a} + \dfrac{1}{2}\right) \qquad (3)$$

The atomic and the external ELF fields interact via the Hamiltonian

$$H_{atom-field} = \hbar g \int E^{(-)}(\hat{x},t)\psi_g^+(\vec{r})\psi_e(\vec{r})d^3\vec{r} + \hbar g \int E^{(+)}(\hat{x},t)\psi_e^+(\vec{r})\psi_g(\vec{r})d^3\vec{r} \qquad (4)$$

Here $g$ is the atom-field coupling constant. $E^{(-)}(\hat{x},t)$ is the negative frequency part of the external ELF electromagnetic field given by

$$E^{(-)}(\hat{x},t) = E_0 \exp[i(\omega_L t - k_L \hat{x} + \phi)] \qquad (5)$$

$E_0$ is the amplitude of the ELF field, $k_L = 2\pi/\lambda_L$ is the wave vector of the filed, and $\phi$ is just some phase.

Finally, the collision Hamiltonian is taken to be

$$H_{atom-atom} = \frac{2\pi\hbar^2 a_{gg}}{m}\int d^3\vec{r}\,\psi_g^+(\vec{r})\psi_g^+(\vec{r})\psi_g(\vec{r})\psi_g(\vec{r})$$

$$\frac{2\pi\hbar^2 a_{ee}}{m}\int d^3\vec{r}\,\psi_e^+(\vec{r})\psi_e^+(\vec{r})\psi_e(\vec{r})\psi_e(\vec{r}) + \frac{4\pi\hbar^2 a_{ge}}{m}\int d^3\vec{r}\,\psi_g^+(\vec{r})\psi_g(\vec{r})\psi_e^+(\vec{r})\psi_e(\vec{r})$$

(6)

Here $a_{ij}$ is the s-wave scattering length in the states $i, j$. Here we have assumed that $a_{eg} = a_{ge}$. The study of the quantum statistical properties of the condensate (at T=0) can be reduced to a relatively simple model by using a mode expansion and subsequent truncation to just a single mode (the "condensate mode"). In particular, one writes the Heisenberg ground state atomic field annihilation operator as a mode expansion over single particle states,

$$\psi_g(\vec{r},t) = \sum_\alpha b_\alpha(t)\phi_\alpha^g(\vec{r})\exp(-i\mu_\alpha t/\hbar) = b_0(t)\phi_0^g(\vec{r})\exp(-i\mu_0 t/\hbar) + \tilde{\psi}(\vec{r},t) \quad (7)$$

Where $\{\psi_\alpha(\vec{r})\}$ are a complete orthonormal basis set and $\{\mu_\alpha\}$ the corresponding eigenvalues. The first term in the second line of eqn.(6) acts only on the condensate state vector, with $\phi_0^g(\vec{r})$ chosen as a solution of the stationary Gross-Pitaevskii equation. The second term, $\tilde{\psi}(\vec{r},t)$, accounts for non-condensate atoms. In a similar manner, we can expand the excited state atomic field annihilation operator

$$\psi_e(\vec{r},t) = \sum_\beta c_\beta(t)\phi_\beta^\alpha(\vec{r})\exp(-i\nu_\beta t/\hbar) \qquad (8)$$

Retaining only the lowest mode of the ground state and the excited state, the Hamiltonian of Eq (1) is written ignoring some constant as energy terms

$$H = \hbar\nu\left(\hat{a}^+\hat{a} + \frac{1}{2}\right) + \frac{\hbar\tilde{\omega}_0}{2}\hat{\sigma}_z + \hbar G\left(E^{(-)}(\hat{x},t)\hat{\sigma}_- + E^{(+)}(\hat{x},t)\hat{\sigma}_+\right) \qquad (9)$$

Where

$$\hbar\omega_b = \int d^3\vec{r}\,\phi_0^{*g}(\vec{r})\left(-\frac{\hbar^2\vec{\nabla}^2}{2m} + V_g(\vec{r})\right)\phi_0^g(\vec{r})$$

$$\hbar\omega_e = \int d^3\vec{r}\,\phi_0^{*e}(\vec{r})\left(-\frac{\hbar^2\vec{\nabla}^2}{2m} + V_e(\vec{r})\right)\phi_0^e(\vec{r})$$

$$\hbar \kappa_{gg} = \frac{4\pi \hbar^2 a_{gg}}{m} \int d^3\vec{r} |\phi_0^g(\vec{r})|^4$$

$$\hbar \kappa_{ee} = \frac{4\pi \hbar^2 a_{ee}}{m} \int d^3\vec{r} |\phi_0^e(\vec{r})|^4$$

$$\hbar \kappa_{eg} = \frac{4\pi \hbar^2 a_{eg}}{m} \int d^3\vec{r} |\phi_0^e(\vec{r})|^2 |\phi_0^g(\vec{r})|^2$$

$$\hbar G = \hbar g \int \phi_0^g(\vec{r}) \phi_0^{*e}(\vec{r}) d^3\vec{r} = \hbar g \int \phi_0^e(\vec{r}) \phi_0^{*g}(\vec{r}) d^3\vec{r}$$

$\hbar \tilde{\omega}_0 = \hbar(\omega_c - \omega_b) + \hbar \gamma_0$, $\hat{\sigma}_{b_0 b_0} + \hat{\sigma}_{c_0 c_0} = 1$, $\hat{\sigma}_{b_0 b_0} = b_0^+ b_0$, $\hat{\sigma}_{c_0 c_0} = c_0^+ c_0$, $\hat{\sigma}_z = \hat{\sigma}_{c_0 c_0} - \hat{\sigma}_{b_0 b_0}$,
$\hat{\sigma}_+ = b_0 c_0^+ \exp(i\tilde{\omega}_0 t)$, $\hat{\sigma}_- = c_0 b_0^+ \exp(-i\tilde{\omega}_0 t)$, $\gamma_0 = (\kappa_{gg} - \kappa_{ee})/2$.

Here both $g$ and G are assumed to be real without loss of generality. The eigenvalues $\mu_0$ and $\nu_0$ are identified as $\mu_0 = \hbar \omega_b - \kappa_{gg}/2$ and $\nu_0 = \hbar \omega_c - \kappa_{ee}/2$ i.e the energy levels shifted by atomic collisions. $\hat{\sigma}_\pm$ and $\hat{\sigma}_z$ satisfy the spin-1/2 algebra of the Pauli matrices,

$$[\hat{\sigma}_-, \hat{\sigma}_+] = -\hat{\sigma}_z, \quad [\hat{\sigma}_-, \hat{\sigma}_z] = 2\hat{\sigma}_- \tag{10}$$

Figure 2 demonstrates the effect of two body atomic collisions on the energy level separation between the ground and the first excited topological coherent mode. When $\kappa_{gg} > \kappa_{ee}$ (i.e. $\gamma_0 > 0$), then the separation increases as compared to the original unperturbed level separation On the other hand when $\kappa_{gg} < \kappa_{ee}$ (i.e. $\gamma_0 < 0$), the energy levels come closer.

We now quantize the position operator $\hat{x}$ in the usual way as

$$\hat{x} = \sqrt{\frac{\hbar}{2\nu m}} (\hat{a} + \hat{a}^+) \tag{11}$$

Substituting into eq.(5) yields

$$E^{(-)} = E_0 \exp i(\phi + \omega_L t) \exp -i\eta(\hat{a} + \hat{a}^+) \tag{12}$$

Where $\eta = \kappa_L \sqrt{\hbar/2vm}$. $\eta$ is typically $10^{-12}$ for the frequency range we are interested.

With $\hat{U} = \exp(-iH_0 t/\hbar)$, $H_0 = \hbar\tilde{\omega}_0\hat{\sigma}_z/2 + \hbar v(\hat{a}^+\hat{a} + 1/2)$, we now transform the Hamiltonian of eq.(9) into that in the interaction picture.

$$H_I = \hat{U}^+ H \hat{U} + i\hbar \frac{d\hat{U}^+}{dt}\hat{U} = \hbar G E_0 \begin{cases} \exp(i\phi)\exp(i\omega_L t)\exp(-i\eta)(\hat{a}\exp(ivt) + \hat{a}^+\exp(-ivt))\hat{\sigma}_- \\ \exp(-i\tilde{\omega}_0 t) + \exp(-i\phi)\exp(-i\omega_L t) \\ \exp(i\eta)(\hat{a}^+\exp(-ivt) + \hat{a}\exp(ivt))\hat{\sigma}_+ \exp(i\tilde{\omega}_0 t) \end{cases}$$

(13)

As $\eta$ is small, we can expand $E^{(-)}(\hat{x},t)$ to first order (i.e. the atom field interaction is spatially linear)

$$H_I \approx \hbar G E_0 \exp(i\phi)\hat{\sigma}_-\left[\exp i(\omega_L - \tilde{\omega}_0)t - i\eta(\hat{a}\exp i(\omega_L - \tilde{\omega}_0 + v)t + \hat{a}^+ \exp i(\omega_L - \tilde{\omega}_0 - v)t)\right]$$
$$+\text{h.c} \quad (14)$$

Consider first the case when $\gamma_0 > 0$ such that $\tilde{\omega}_0 - v = \tilde{\Omega}$. Now suppose the external ELF field is tuned to resonance i.e. $\omega_L = \tilde{\omega}_0$. We shall then have

$$H_I \approx \hbar G E_0 \exp(i\phi)\hat{\sigma}_-\left[1 - i\eta(\hat{a}\exp(iv)t + \hat{a}^+ \exp(-iv)t)\right] + \text{h.c} \quad (15)$$

Invoking the rotating wave approximation yields

$$H_I \approx \hbar G E_0\left[1 - i\eta(\hat{\sigma}_- \exp(i\phi) + \hat{\sigma}_+ \exp(-i\phi))\right] \quad (16)$$

The equation of motion for the population inversion operator $\hat{\sigma}_z$ is

$$\ddot{\hat{\sigma}}_z - 4\Omega\hat{\sigma}_z = 0 \quad (17)$$

Where $\Omega = GE_0$ is the Rabi frequency associated with the semi-classical atom-field interaction. Eqn. (17) produces Rabi type oscillations between collective excitation sublevels of the same degree of excitation. In other words the BEC oscillates between the ground and excited topological states without any change in its collective excitation state (no change in its shape). If now, we tune the ELF field as $\omega_L = \tilde{\omega}_0 + v$. We shall then have

$$H_I \approx -i\hbar\eta\Omega\exp(i\phi)\hat{a}^+\hat{\sigma}_- + \text{h.c} \quad (18)$$

This is the Jaynes-Cummings Hamiltonian. This Hamiltonian produces Rabi oscillations between $|g,n\rangle$ and $|e,n-1\rangle$ energy levels. The quanta of collective excitation are $n = \hat{a}^+\hat{a}$. Similarly if $\omega_L = \tilde{\omega}_0 - v$, then the resulting interaction corresponds to an anti-Jaynes Cummings model. By keeping more terms in the expansion of $E^{(-)}(\hat{x},t)$ i.e. retaining the spatially nonlinear terms in the atom-field interaction, other higher order Jaynes-Cummings type interaction can be generated. For example, if we retain terms in second order and set $\omega_L = \tilde{\omega}_0 + 2v$, then we get the two quanta interaction Hamiltonian

$$H_I \approx \eta^2 \hat{a}^{+2} \hat{\sigma}_- + \text{h.c} \tag{19}$$

Consider now the case when $\gamma_0 < 0$, so that the level spacing between $|g\rangle$ and $|e\rangle$ decreases such that $\tilde{\omega}_0 = v$. Under such condition, even in the absence of the ELF electromagnetic field ($\omega_L = 0$), an anti Jaynes-Cummings type of interaction is produced. If the Elf field is tuned as $\omega_L = v$ eqn. (16) is reproduced. If the level spacing between the topological modes is further reduced such that $\tilde{\omega}_0 < v$, $v - \tilde{\omega}_0 = \tilde{\Omega}$ then a Jaynes-Cummings type of dynamics is produced when $\omega_L = 2v - \tilde{\Omega}$.

**Conclusion**

In conclusion, we have developed a theory of resonance formation of coupled coherent topological–collective modes of a Bose-Einstein condensate. In order to generate this coupling, the frequency of the extremely low frequency electromagnetic field has to be appropriately tuned to resonance. This tuning depends on the trap frequency and the energy level separation between the ground and the first excited topological coherent mode modified by the two body atomic collisions. We have also showed that the coupling between topological and collective modes can be linear as well as non-linear.

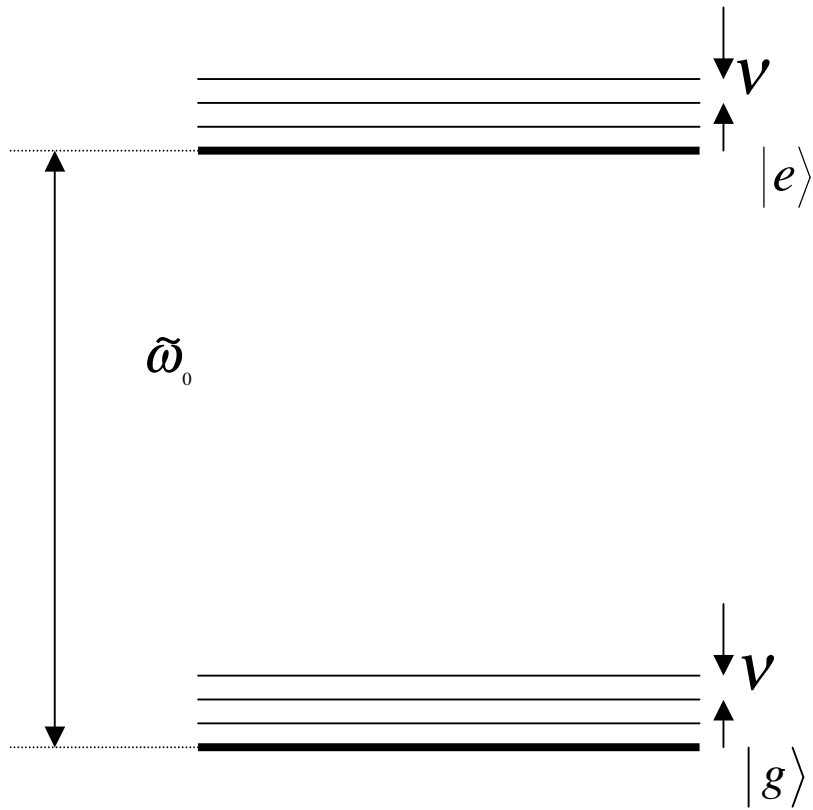

Figure 1: The topological coherent levels $|g\rangle$ and $|e\rangle$ are separated by $\hbar\tilde{\omega}_0$. Within these topological levels are the energy levels of the collective elementary excitations separated by $\hbar\nu$. The unperturbed energy level separation $\hbar\omega_0$ is modified by the two body atomic collisions.

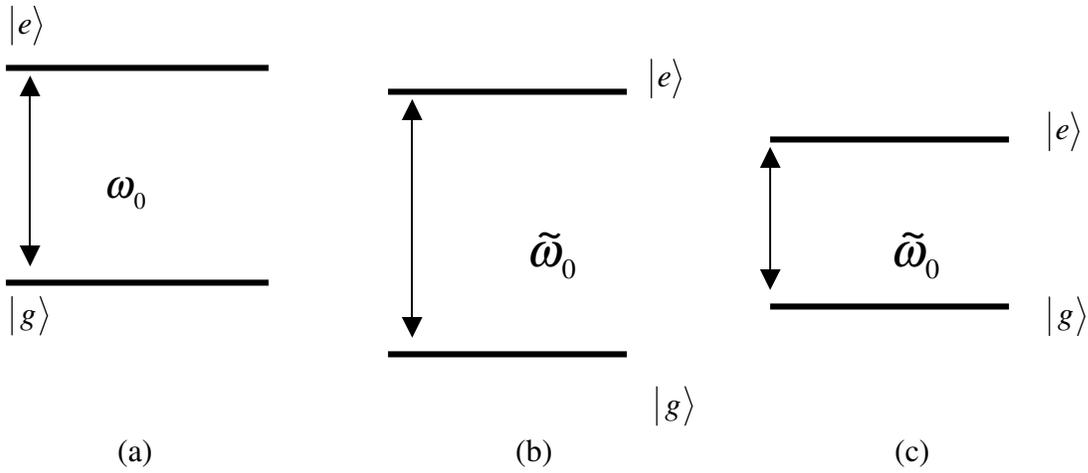

Figure 2.The effect of two body atomic collisions on the energy level separation between the ground and the first excited topological coherent mode is shown. The original unperturbed level is shown in (a). When $\kappa_{gg} > \kappa_{ee}$ (i.e. $\gamma_0 > 0$), then the separation increases as compared to the original unperturbed level separation (b). On the other hand when $\kappa_{gg} < \kappa_{ee}$ (i.e. $\gamma_0 < 0$), the energy levels come closer (c).